\documentclass[conference]{IEEEtran}
\IEEEoverridecommandlockouts
\usepackage{cite}
\usepackage{amsmath,amssymb,amsfonts}
\usepackage{algorithmic}
\usepackage{graphicx}
\usepackage{colortbl}
\usepackage{textcomp}
\usepackage{graphicx}
\usepackage{multirow}
\usepackage{tikz}
\def\BibTeX{{\rm B\kern-.05em{\sc i\kern-.025em b}\kern-.08em
    T\kern-.1667em\lower.7ex\hbox{E}\kern-.125emX}}
\begin{document}

\title{Privacy Enhanced Speech Emotion Communication using Deep Learning Aided Edge Computing
\thanks{Email: siddique.latif@usq.edu.au.}
}

\author{\IEEEauthorblockN{Hafiz Shehbaz Ali}
\IEEEauthorblockA{\textit{EmulationAI} \\
Brisbane, Australia}
\and
\IEEEauthorblockN{Fakhar ul Hassan}
\IEEEauthorblockA{\textit{Information Technology University (ITU)} \\
 Lahore, Punjab, Pakistan}
\and
\IEEEauthorblockN{Siddique Latif}
\IEEEauthorblockA{\textit{University of Southern Queensland} \\
Australia}
\and
\IEEEauthorblockN{Habib Ullah Manzoor}
\IEEEauthorblockA{\textit{University of Engineering and Technology} \\
Lahore, Pakistan}
\and
\IEEEauthorblockN{Junaid Qadir}
\IEEEauthorblockA{\textit{Information Technology University (ITU)} \\
Lahore, Pakistan.}
}

\maketitle

\begin{abstract}
Speech emotion sensing in communication networks has a wide range of applications in real life. In these applications, voice data are transmitted from the user to the central server for storage, processing, and decision making. However, speech data contain vulnerable information that can be used maliciously without the user’s consent by an eavesdropping adversary. In this work, we present a privacy-enhanced emotion communication system for preserving the user personal information in emotion-sensing applications. We propose the use of an adversarial learning framework that can be deployed at the edge to unlearn the users' private information in the speech representations. These privacy-enhanced representations can be transmitted to the central server for decision making. We evaluate the proposed model on multiple speech emotion datasets and show that the proposed model can hide users' specific demographic information and improve the robustness of emotion identification without significantly impacting performance. To the best of our knowledge, this is the first work on a privacy-preserving framework for emotion sensing in the communication network.    
\end{abstract}


\section{Introduction}
\label{sec:introduction}
Nowadays, the integration of the internet of things (IoT) and artificial intelligence (AI) has made great progress and created various real-world applications \cite{latif20175g,latif2017artificial}. Most importantly, human-computer interaction (HCI) systems-based solutions such as speech assistants, service and social robots are becoming an integral part of our life \cite{latif2020speech,latif2021survey}. Among them emotional sensing systems have become a hot area of research due to their many applications in call centres \cite{burkhardt2006detecting}, forensic sciences \cite{roberts2012forensic}, smart cars \cite{vogel2018emotion}, and healthcare \cite{latif2020speech}. Emotion recognition systems have achieved significant outcomes using voice, facial expression, or physiological signals. In this paper, we focus on speech emotion recognition (SER), which has become more popular due to its wide range of applications and availability of voice data.

\begin{figure}[!ht]
\centering
%captionsetup{justification=centering}
\includegraphics[trim=0cm 0cm 0cm 0cm,clip=true,width=0.4\textwidth]{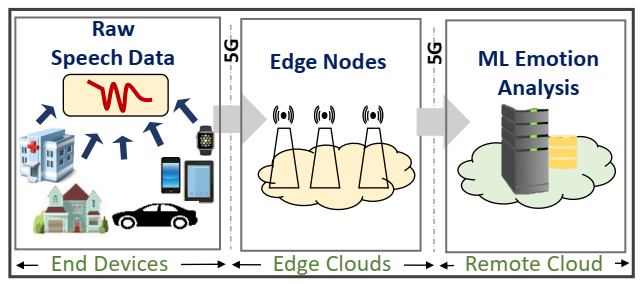}
%\captionsetup{width=0.95\textwidth}
\caption{The state-of-the-art emotion-sensing system that transmits raw speech signal over the communication network for emotion analysis.}
\label{fig:state}
\end{figure}

Existing studies on voice-based emotion detection mainly focusing on improving the accuracy of the systems for enabling their real-time applications \cite{latif2020deep,latif2020deep1}. In emotion-sensing applications, deep learning (DL), IoT, and communication technologies are playing a prime role to provide services at various scales. Most of the emotion-sensing services follow the system model in which raw speech is transmitted to the remote server for processing and decision making. This has been shown in Figure \ref{fig:state}. Such systems are successful in real-life, however, they involve complete sharing of speech over the communication network, which may lead to adverse consequences to people's privacy \cite{jaiswal2020privacy}. Speech signal contains sensitive information about the message, speaker, gender, language, etc., which may be misused by eavesdropping adversary without users' consent \cite{latif2020federated}. %Research have shown that DL based applications in communication systems are vulnerable to adversarial attacks. 

% %In all these attempts, recent advancements in deep learning (DL) are playing prime role in the development of new systems with better performance.

% These days, adversarial learning based systems are becoming more popular for speech emotion recognition due to the powerful feature modelling abilities \cite{latif2020deep}. However, the research on real-time deployment of \textcolor{red}{speech emotion recognition} (SER) systems is a hot area of research. Some studies also proposed different frameworks by utilisation of IoT and communication technologies for different applications of SER systems in daily life.

In this work, we propose a privacy-preserving representation learning module that can be deployed at the edge servers to enhance the privacy of users in SER applications. We propose to use an adversarial learning framework at the edge nodes to unlearn the specific demographic information in speech representations and encode the data in the compressed form prior to offloading over the communication network. In this way, proposed system communicate privacy-enhanced features with the central server, which masks our proposed framework more secure and suitable for real-life different applications including hybrid driving, healthcare, voice-based social assistants, and many more. We evaluate the proposed framework on different publicly available datasets and show that how the users' privacy can be improved in terms of demographic information (speaker identity, gender information, and spoken language) without significantly impacting the performance of emotion recognition. To best of our knowledge, this is a first attempt to achieve secure emotion communication using DL and edge computing.

The rest of the paper is organised as follows. Section \ref{Sec:rel} presents recent related work on speech emotion communication. Section \ref{sec:healthcareOpportunities} explains the proposed framework of privacy-enhanced speech emotion communication. Section \ref{experi} covers experimental setup and Section \ref{resu} presents results and discussions. Finally, Section \ref{con} concludes the paper and discusses future research directions.

\section{Related Work}
\label{Sec:rel}

Previous works on SER have achieved significantly improved results using DL techniques. Researchers also proposed different framework based on advanced communication technologies for real-time deployments. For instance, Hossain et al. \cite{hossain2017emotion} introduced an IoT-based framework for emotion-aware connected communication systems. The framework captures and processes speech and images signals to classify patient emotions. The signal processing is distributed among cloudlets, the edge cloud, and the remote cloud. The results based on both publicly available and locally collected datasets showed that the proposed system can be successfully deployed in any 5G emotion-aware healthcare framework. However, this work was not focused on privacy issues. Chen et al. \cite{chen20175g} introduced a 5G-based cognitive framework for healthcare big data analytics. The framework constructs resource cognitive and data cognitive engines using soft-denied networks with AI support The proposed system provides reliable support for a better realisation of emotion communication over human to human and human to machine networks. Zhou et al. \cite{zhou2019multi} proposed a multi-task communication system to solve resource allocation issues. The aforementioned scheme designed a transmission frame structure for emotion data that allows real-time multiple service requests and improve the resource utilisation ratio. In another work, Li et al. \cite{8933554} proposed an AI-based emotion communication system to cast emotion data efficiently in the network communication. The system is evaluated under two application scenarios including unmanned driving and emotional social robots. In this framework, data processing and labelling are performed at the edge cloud which emotion classification decisions are made at the remote cloud. % However, the privacy of users and system in the context of emotion data has not been considered in these studies. 

Most of the above-mentioned studies considered to transmit raw speech signal to remote for processing and decisions making. Speech signal also contains demographic information that can be used for the identification of users. Some recent studies \cite{jaiswal2020privacy,aloufi2020paralinguistic} have shown that the attacker can use such information from speech signal to identify speaker and gender information. In this paper, we are presenting an emotion communication system based on DL and edge computing for users' privacy protection in emotion-sensing applications.

%Privacy preservation of speech data is of a great importance and posses several challenges for network scientists and healthcare experts. In order to deal with these challenges, since last decade several works have been proposed and reported.

% \begin{figure*}[t]
%     \centering
%     \captionsetup{justification=centering}
%     \subfloat[(a)]{\includegraphics[clip,trim=4cm 8cm 4cm 3cm, scale=0.66]{pictures/ProposedModel4aa.pdf}}%
%     \qquad
%     \subfloat[(b)]{\includegraphics[clip,trim=4cm 8cm 4cm 3cm, scale=0.66]{pictures/ProposedModel4bb.pdf} }%
%     \caption{Proposed model.}%
%     \label{ProposedModel}%
% \end{figure*}

\section{Proposed Privacy Enhanced Emotion Communication}
\label{sec:healthcareOpportunities}

\begin{figure}[!ht]
\centering
%captionsetup{justification=centering}
\includegraphics[trim=0cm 0cm 0cm 0cm,clip=true,width=0.4\textwidth]{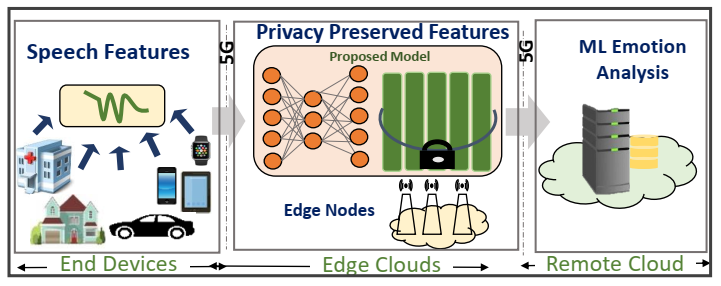}
%\captionsetup{width=0.95\textwidth}
\caption{The proposed emotion-sensing system that transmits privacy-enhanced features over the communication network. An adversarial model is being used to encode input features to privacy-enhanced representations. }
\label{fig:proposed}
\end{figure}

Due to the rise in mobile devices, privacy is very crucial for real-life applications \cite{latif2017mobile}. In particular, SER application users provide complete access to their voice recordings for speech analytics. The undesired access to users' speech can be used for various malicious ways. This motivates us to design the privacy-enhanced emotion communication system that uses edge computing and adversarial machine learning. Figure \ref{fig:proposed} illustrates an overview of the proposed architecture. The proposed architecture comprises of \textit{speech sensing layer}, \textit{edge computing layer}, and \textit{emotion analysis \& decision making layer}. In the following, we present the details of each
component.

\subsection{Speech Sensing Layer}
% Speech sensing layer consists of end user devices, which collects speech data from cyber-physical space. Instead of transmitting raw speech to edge servers, it convert them into speech features and request service from the edge servers. 

The speech sensing layer consists of end-user devices, which collect speech data from cyber-physical space. Instead of transmitting raw speech to edge servers, this layer processes it and converts it into the speech features. These speech features are propagated to the edge servers for privacy-enhanced feature generation.

\subsection{Edge Computing Layer}
\begin{figure}[!ht]
\centering
%captionsetup{justification=centering}
\includegraphics[trim=0cm 0cm 0cm 0cm,clip=true, width=0.38\textwidth]{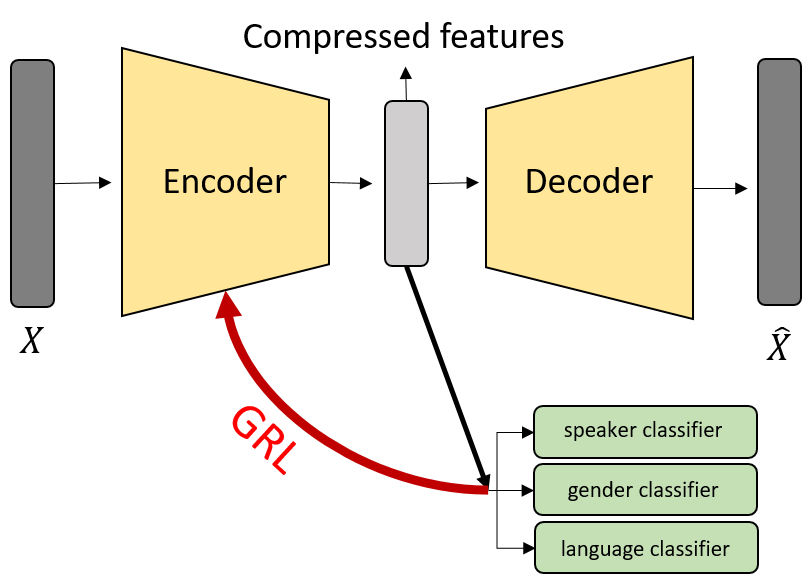}
%\captionsetup{width=0.95\textwidth}
\caption{Block diagram of the proposed multi-task learning-based autoencoder model that use GRL for privacy enhanced feature extraction.  }
\label{fig:model}
\end{figure}

The edge server is an important component of our proposed architecture. The edge server leverages low-consumption computational and storage hardware such i.e., edge cloudlets \cite{taleb2017multi} and operates within a radio access network (RAN) in the close vicinity of end-users \cite{liu2020toward}. In our proposed architecture (as shown in Figure \ref{fig:proposed}), the role of the edge server is not limited to traffic aggregation gateway and network service controller only. It acts as an intelligent edge server that is also responsible to minimise the dimensions of speech features and represent them into privacy-preserving representation while retaining the maximum data utility as much as possible. To achieve this, we propose to deploy an adversarial network (see Figure \ref{fig:model}) at the edge nodes. These representations then transmitted to remote cloud servers for emotion identification and decision-making tasks. This data dimensional reduction functionality also reduces cloud (core) server data processing and analytics loads by transmitting only important speech features. In short, the edge computing layer reduces data dimensions and controls the exchange of data between end devices and the remote cloud server. 

% It acts as mini cloud server that is responsible to minimise the dimension of speech features and represent them into privacy preserving representation while maintaining the data utility as much as possible. These representations than transmitted to cloud servers for emotion identification and decision making for different related applications. In other words, edge computing layer controls the exchange of data between voice recording devices and the remote cloud server. 

% trim={<left> <lower> <right> <upper>}

\textbf{Privacy Preserved Feature Extraction:} Our proposed model is built based on a multi-task learning setting which includes an autoencoder network and three classifiers for gender, speaker and language classification (see Figure \ref{fig:model}). The autoencoder network consists of encoder and decoder networks where the encoder is responsible for encoding the input features to latent representation and decoder use the latent representations to reconstruct the features. Latent representations of given speech include emotional state, the content of speech and speaker’s characteristics such as age, gender, accent, language and speaker traits. These latent variables are used by machine learning models for emotion detection. Here, we aim to hide the sensitive information about speaker identity, gender and language. Therefore, we adversarially connected three classifiers to the autoencoder network that use latent representations and use gradient reversal (GRL) \cite{ganin2015unsupervised} to unlearn demographic information about the user, gender, and spoken language. Here, we use AE to learn emotional representations. For a given input data $X$, AE objective function can be defined as:
\begin{equation}
\label{AE}
    \mathcal{L_{\text{AE}}}(x,D_{\phi_{e}}(E_{\theta_{e}}(x)))=\lVert{x-\hat{x}}\rVert_{2}^{2}.
\end{equation}
Given the encoder parameters $\theta_{e}$, the adversarial branch minimises $\theta_{s}$, $\theta_{g}$, and $\theta_{l}$ for three classifiers: speaker, gender, and language. Overall, the model is trained in an end-to-end manner by optimising the following min-max objective:
\begin{equation}
\begin{aligned}
\label{K1}
    \underset{\theta_{e},\phi_{e}}{\text{min}} \quad\underset{\theta_{s},\theta_{g},\theta_{l}}{\text{max}}\quad\mathcal{L}_{\text{AE}}(\theta_e, \phi_e; X) \quad \quad \quad \quad \quad \quad \quad \quad  \\ -\alpha \Big[\mathcal{L}_{\text{spk}}(\theta_{e},\theta_{s})+\mathcal{L}_{\text{ge}}(\theta_{e},\theta_{g})+\mathcal{L}_{\text{lan}}(\theta_{e},\theta_{l})\Big]
\end{aligned}
\end{equation}
where $\alpha$ is the trade-off parameter. The baseline network is considered by setting $\alpha=0$. The max part in equation \ref{K1} corresponds to the adversary which only control the speaker, gender, and language parameters $\theta_{s}$, $\theta_{g}$, and $\theta_{l}$ respectively. %\textcolor{red}{ Algorithm \textbf{1} explains the step by step training of the proposed multi-task learning based deep autoencoder network for speech feature selection at edge layer.  }

\subsection{Emotion Analysis \& Decision making Layer}
The cloud server is responsible for emotion analytics, decision making, and storage services. We deploy the emotion classifier in a cloud server that uses privacy-enhanced speech representations and performs emotion identification tasks. In this way, an emotion communication system provides secure services by filtering sensitive information at the edge nodes.  

%\SK {Sid bhai: Which Classifier is used, in the results SVM is mentioned. Please mention here SVM as well}
\begin{table*}[!ht]
\centering
\caption{Corpora information and the mapping of categorical emotions onto Negative/Positive valence.}
\begin{tabular}{|l|l|l|l|l|}
\hline
Corpus                                         & Language                                      & Speakers                                      & Negative Valance                              & Positive Valance                              \\ \hline
\begin{tabular}[c]{@{}l@{}}IEMOCAP\end{tabular} & \begin{tabular}[c]{@{}l@{}}English\end{tabular} & \begin{tabular}[c]{@{}l@{}}10\end{tabular} & \begin{tabular}[c]{@{}l@{}}Angry, Sadness\end{tabular} & \begin{tabular}[c]{@{}l@{}}Neutral, Happy, Excited\end{tabular} \\ \hline

\begin{tabular}[c]{@{}l@{}}EMODB\end{tabular}  & \begin{tabular}[c]{@{}l@{}}German\end{tabular} & \begin{tabular}[c]{@{}l@{}}10\end{tabular} & \begin{tabular}[c]{@{}l@{}}Anger, Sadness, Fear, Disgust, Boredom\end{tabular} & \begin{tabular}[c]{@{}l@{}}Neutral, Happiness\end{tabular} \\ \hline
\begin{tabular}[c]{@{}l@{}}BUEMODB\end{tabular}  & \begin{tabular}[c]{@{}l@{}}Turkish\end{tabular} & \begin{tabular}[c]{@{}l@{}}11\end{tabular} & \begin{tabular}[c]{@{}l@{}}Anger, Sadness\end{tabular} & \begin{tabular}[c]{@{}l@{}}Neutral, Joy\end{tabular} \\ \hline

\begin{tabular}[c]{@{}l@{}}EMOVO\end{tabular}  & \begin{tabular}[c]{@{}l@{}}Italian\end{tabular} & \begin{tabular}[c]{@{}l@{}}4\end{tabular} & \begin{tabular}[c]{@{}l@{}}Anger, Sadness, Fear, Disgust\end{tabular} & \begin{tabular}[c]{@{}l@{}}Neutral, Joy, Surprise \end{tabular} \\ \hline
\end{tabular}
\label{data}
\end{table*}

% \SetCommentSty{mycommfont}

% \begin{algorithm}[h]
% \DontPrintSemicolon
% \textbf{Procedure}  \Procedure{AETrain}{($X$,$\alpha$,$\theta_{s}$, $\theta_{g}$, $\theta_{l}$)}\\
% \textbf{Input:} Speech feature data $X$, speaker, gender, and language parameters [$\theta_{s}$, $\theta_{g}$,
% $\theta_{l}$], batch size, number of epoch $\zeta$, and learning rate\\
% \textbf{Output:} New feature representations  $X^'$, \\
% \tcc{Training of Encoder and Decoder Network}\\
% Initialization; \\
% Set autoencoder (AE) model parameters;\\
% \For{($j$ $\in$ $\zeta$)}{\\
% %Train Encoder and Decoder network using Eq. \ref{AE}\\
% Optimise model using Eq. \ref{K1}.}\\
% End training and calculate UAR score
% \textcolor{red}{\caption{Autoencoder Model Training}}
% \label{AETrain}
% \end{algorithm}

\section{Experimental Setup}
\label{experi}

\subsection{Datasets}
In order to evaluate the performance of proposed model, we use different publicly available datasets including IEMOCAP \cite{busso2008iemocap}, EMODB \cite{burkhardt2005database}, EMOVO \cite{costantini2014emovo}, and BUEMODB \cite{kaya2016robust}. 
The selection of these datasets is made to rigorously evaluate the proposed algorithm for gender, speaker, and language. All these corpora are annotated differently, therefore, we follow a consistent way as used in the previous studies \cite{latif2019unsupervised,latif2018transfer} to evaluate the proposed framework by mapping their emotional labels to binary positive/negative valence. Emotional mapping to binary valence is given in Table \ref{data}. Further details of these datasets are presented below.  \linebreak
\begin{enumerate}
    \item \textbf{IEMOCAP} is a multimodal emotional English corpus, which contains dyadic conversations of 10 professional actors (five female and five male). Overall data consists of five sessions and utterances are annotated in 10 emotions. To be consistent with previous studies \cite{latif2018variational}, we use 4 emotions classes including angry, happy, neutral and sad. 
    \item \textbf{EMODB} is an emotional speech database in the German language. Ten actors (5 female and 5 male) simulated 7 emotions including neutral, anger, fear,  joy, sadness, disgust, and boredom. EMODB corpus contains German utterances of 10 sentences from everyday communication. We use all emotional utterances in this work.
    \item \textbf{EMOVO} is the first emotional corpus in the Italian language. It contains voices of 6 actors (three males and three females) in seven emotional states (disgust, fear, anger, joy, surprise, sadness, neutral). The actors simulated emotional states in 14 sentences. 
    \item \textbf{BUEMODB} is a Turkish language corpus recorded by 11 amateur actors (7 female and 4 male) simulating 11 emotionally undefined sentences. Recordings are labelled in four basic emotions including joy, neutral, anger, and sadness.
\end{enumerate}
% \textbf{IEMOCAP} is a multimodal emotional English corpus, which contains dyadic conversations of 10 professional actors (five female and five male ). Overall data consists of five sessions and utterances are annotated in 10 emotions. To be consistent with previous studies \cite{latif2018variational}, we use 4 emotions classes including angry, happy, neutral and sad. \linebreak
% \textbf{EMODB} is a emotional speech database in German language. Ten actors (5 female and 5 male) simulated 7 emotions including neutral, anger, fear,  joy, sadness, disgust, and boredom. EMODB corpus contains German utterances of 10 sentences from everyday communication. We use all emotional utterances in this work. \linebreak
% \textbf{EMOVO} is first emotional corpus in Italian language. It contains voices of 6 actors (three males and three females) in seven emotional states (disgust, fear, anger, joy, surprise, sadness, neutral). The actors simulated emotional states in 14 sentences. \linebreak
% \textbf{BUEMODB} is Turkish language corpus recorded by 11 amateur actors (7 female and 4 male) simulating 11 emotionally undefined sentences. Recording are labelled in four basic emotions including joy, neutral, anger, and sadness.

\subsection{Speech Features}
In this work, we used Interspeech Computational Paralinguistics Challenge features set (COMPARE) \cite{schuller2013interspeech,schuller2014interspeech} for speech representation. The COMPARE feature set includes $6373$ static features resulting from the computation of various functionals over low-level descriptor (LLD) contours. Complete detail about the feature set can be found in \cite{weninger2013acoustics}. We use openSMILE toolkit \cite{eyben2013recent} for extracting these features from speech utterances.

\subsection{Model Configuration}
We implemented our model with feed-forward neural network layers. The encoder and decoder network contains two hidden layers of hidden units of 2\,000 and 1000 each. We set the dimension of latent features to 512. We use the dropout layer between two consecutive layers of AE for regularisation. We use the dropout value of 0.5. We select Leaky Rectified Linear Units (leaky ReLUs) \cite{xu2015empirical} as activation function for all hidden layers. For gender, speaker, and language classifications, we use two dense layers of 400 hidden units. We also use the dropout layer between two dense layers of classification networks with a dropout value of 0.5. We use a learning rate of $10^{-5}$ in all of our experiments. 

For emotion classification, we follow leave-on-speaker-out scheme \cite{latif2019unsupervised} and trained support vector machines (SVMs) on the encoded features by our privacy-enhanced features extraction module. We select an RBF kernel and performed a grid search using validation data to pick the optimal hyperparameters. We used the unweighted average recall (UAR) as the performance metric. We repeat all experiments five times and mean results are reported. We apply min-max normalisation to the input features to the model. 

\section{Results and Discussions}
\label{resu}
The objective of this proposed system in to maximise the privacy of speech emotion communication using the edge computing. We proposed the use of privacy preserved feature extraction model in edge computing layer. In this section, we evaluate the proposed architecture against different scenarios where an attacker could maliciously attack the system to obtain sensitive demographic information. 

\subsection{Attacker Network}
We assume that the attacker has access to the testing dataset or portion of test data with known language and gender labels. The attacker uses the main network to generate representations for this dataset and use its own network to predict language and gender information. For the attacker network, we use two dense layers to compute the results for prediction of gender and language labels. For the speaker recognition task, we follow the study \cite{jaiswal2019privacy} and consider user identification which is the possibility of an attacker to identify either a person is a member of the system or belong to the training set.

\subsection{Evaluations}
We performed emotion classification on the privacy-enhanced features encoded by our proposed adversarial model. We fed COMPARE features to the privacy-enhanced feature extraction module of the proposed framework and encode them to the compressed form of 512 features. These encoded features are used for emotion, gender and user identification. Results are shown in Table \ref{results}. We also implemented Principal component analysis (PCA) \cite{pearson1901liii} and standard autoencoder (AE) network to compare the results with our proposed model. We set $\alpha =0$ in Equation \ref{K1} for baseline AE implementation. 
\begin{table}[!ht]
\centering
\caption{Results comparison using standard and privacy preserving features with different methods. }
\begin{tabular}{|l|l|l|l|l|}

\hline
Method                 & emotion     & gender     & user     & language    \\ \hline
\multicolumn{5}{|c|}{Performance on standard features}                    \\ \hline
COMPARE               & 71.5       & 90.5       & 68.2 & 78.2           \\ \hline
Autoencoder                    & 69.5        &  85.3    &  65.2  & 72.5           \\ \hline
PCA           & 66.5        &  84.6      & 64.5  & 71.2             \\ \hline
\multicolumn{5}{|c|}{\textbf{Performance on privacy preserving features}} \\ \hline
\rowcolor[HTML]{EFEFEF} 
\textbf{Proposed}      & 68.7       &  71.2      & 54.1 &  60.1           \\ \hline
\end{tabular}
\label{results}
\end{table}
Results are presented in Table \ref{results}, which shows the performance of standard and privacy-enhanced features. It can be noted that unweighted average recall (UAR) of emotion classification using COMPARE features is 71.5 \% with very high accuracy (e.g., above 90 \% on gender identification) on all other tasks, i.e., gender, language, user identification. This shows the emotional features, i.e., COMPARE, also contain the highly discriminating information about the speaker, gender, and language identification. This shows that emotional features cannot be transmitted to the network as they can leak demographic information over the network. Emotional representations are also discriminating for speaker, gender, and language identification even compressing the COMPARE features using autoencoder and PCA. In contrast, our proposed model can extract emotional features by preserving the information about the user, gender, and language. It can be noted in Table \ref{results} that the proposed model can improve robustness by minimising the performance of the speaker, gender, and language identification. 

\begin{figure}[!ht]
\centering
%captionsetup{justification=centering}
\includegraphics[trim=0cm 0.6cm 0cm 0cm,clip=true, width=0.4\textwidth]{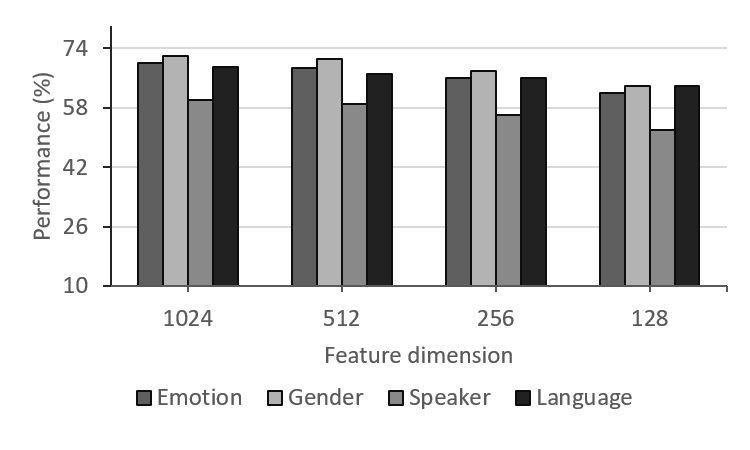}
%\captionsetup{width=0.95\textwidth}
\caption{Effect of changing the dimension of privacy enhanced features on the performance of the system.}
\label{fig:res}
\end{figure}

We also varied the dimension of latent features and plotted the graph of the performance of the model in Figure \ref{fig:res}. This shows that small dimensions of features provide more robustness in terms of the speaker, language, and gender identification with minor degradation in SER performance. 

% trim={<left> <lower> <right> <upper>}

\section{Conclusions} 
\label{con}

%\SL{Sana please write this}

We proposed a novel deep learning-based privacy-enhanced representation learning architecture for emotion-sensing applications. The proposed framework leverages edge nodes to encode speech features to robust and compressed representation to be transmitted to the network. We use adversarial learning framework for privacy-preserving features learning by unlearning the information about the speaker, gender and language representation. This makes the proposed framework suitable for real-life applications including hybrid driving, healthcare, and voice-based social assistants. We evaluated the proposed framework on four publicly available datasets. Based on the results, we show that the proposed framework enhances the privacy of the emotion-sensing system in terms of speaker, gender, and language without significantly degrading the emotion identification performance. In future, we aim to study the gains of deep-learning-based speech emotion communication system under the real edge-based mobile cloud and cloud-let scenarios.

% edge computing enabled $5G$ networks to enhance the privacy of users in speech emotion communication applications. 

% Our proposed adversarial deep learning-based autoconder model encapsulates users' private information from speech data and encode the data in compressed representations prior to offloading to the network. 

% \bibliographystyle{IEEEtran}
% \bibliography{Reference}

\begin{thebibliography}{10}
\providecommand{\url}[1]{#1}
\csname url@samestyle\endcsname
\providecommand{\newblock}{\relax}
\providecommand{\bibinfo}[2]{#2}
\providecommand{\BIBentrySTDinterwordspacing}{\spaceskip=0pt\relax}
\providecommand{\BIBentryALTinterwordstretchfactor}{4}
\providecommand{\BIBentryALTinterwordspacing}{\spaceskip=\fontdimen2\font plus
\BIBentryALTinterwordstretchfactor\fontdimen3\font minus
  \fontdimen4\font\relax}
\providecommand{\BIBforeignlanguage}[2]{{%
\expandafter\ifx\csname l@#1\endcsname\relax
\typeout{** WARNING: IEEEtran.bst: No hyphenation pattern has been}%
\typeout{** loaded for the language `#1'. Using the pattern for}%
\typeout{** the default language instead.}%
\else
\language=\csname l@#1\endcsname
\fi
#2}}
\providecommand{\BIBdecl}{\relax}
\BIBdecl

\bibitem{latif20175g}
S.~Latif, J.~Qadir, S.~Farooq, and M.~Imran, ``How 5g wireless (and concomitant
  technologies) will revolutionize healthcare?'' \emph{Future Internet},
  vol.~9, no.~4, p.~93, 2017.

\bibitem{latif2017artificial}
S.~Latif, F.~Pervez, M.~Usama, and J.~Qadir, ``Artificial intelligence as an
  enabler for cognitive self-organizing future networks,'' \emph{IEEE
  Communications Society (ComSoc)'s blog on Cognitive Radio Networking and
  Security}, 2017.

\bibitem{latif2020speech}
S.~Latif, J.~Qadir, A.~Qayyum, M.~Usama, and S.~Younis, ``Speech technology for
  healthcare: Opportunities, challenges, and state of the art,'' \emph{IEEE
  Reviews in Biomedical Engineering}, 2020.

\bibitem{latif2021survey}
S.~Latif, H.~Cuay{\'a}huitl, F.~Pervez, F.~Shamshad, H.~S. Ali, and E.~Cambria,
  ``A survey on deep reinforcement learning for audio-based applications,''
  \emph{arXiv preprint arXiv:2101.00240}, 2021.

\bibitem{burkhardt2006detecting}
F.~Burkhardt, J.~Ajmera, R.~Englert, J.~Stegmann, and W.~Burleson, ``Detecting
  anger in automated voice portal dialogs,'' in \emph{Ninth International
  Conference on Spoken Language Processing}, 2006.

\bibitem{roberts2012forensic}
L.~S. Roberts, ``A forensic phonetic study of the vocal responses of
  individuals in distress,'' Ph.D. dissertation, University of York, 2012.

\bibitem{vogel2018emotion}
H.-J. V{\"o}gel, C.~S{\"u}{\ss}, T.~Hubregtsen, E.~Andr{\'e}, B.~Schuller,
  J.~H{\"a}rri, J.~Conradt, A.~Adi, A.~Zadorojniy, J.~Terken \emph{et~al.},
  ``Emotion-awareness for intelligent vehicle assistants: A research agenda,''
  in \emph{2018 IEEE/ACM 1st International Workshop on Software Engineering for
  AI in Autonomous Systems (SEFAIAS)}.\hskip 1em plus 0.5em minus 0.4em\relax
  IEEE, 2018, pp. 11--15.

\bibitem{latif2020deep}
S.~Latif, ``Deep representation learning for improving speech emotion
  recognition,'' \emph{Doctoral Consortium, Interspeech 2020}, 2020.

\bibitem{latif2020deep1}
S.~Latif, R.~Rana, S.~Khalifa, R.~Jurdak, J.~Qadir, and B.~W. Schuller, ``Deep
  representation learning in speech processing: Challenges, recent advances,
  and future trends,'' \emph{arXiv preprint arXiv:2001.00378}, 2020.

\bibitem{jaiswal2020privacy}
M.~Jaiswal and E.~M. Provost, ``Privacy enhanced multimodal neural
  representations for emotion recognition.'' in \emph{AAAI}, 2020, pp.
  7985--7993.

\bibitem{latif2020federated}
S.~Latif, S.~Khalifa, R.~Rana, and R.~Jurdak, ``Federated learning for speech
  emotion recognition applications,'' in \emph{2020 19th ACM/IEEE International
  Conference on Information Processing in Sensor Networks (IPSN)}.\hskip 1em
  plus 0.5em minus 0.4em\relax IEEE, 2020, pp. 341--342.

\bibitem{hossain2017emotion}
M.~S. Hossain and G.~Muhammad, ``Emotion-aware connected healthcare big data
  towards 5g,'' \emph{IEEE Internet of Things Journal}, vol.~5, no.~4, pp.
  2399--2406, 2017.

\bibitem{chen20175g}
M.~Chen, J.~Yang, Y.~Hao, S.~Mao, and K.~Hwang, ``A 5g cognitive system for
  healthcare,'' \emph{Big Data and Cognitive Computing}, vol.~1, no.~1, p.~2,
  2017.

\bibitem{zhou2019multi}
P.~Zhou, M.~S. Hossain, X.~Zong, G.~Muhammad, S.~U. Amin, and I.~Humar,
  ``Multi-task emotion communication system with dynamic resource
  allocations,'' \emph{Information Fusion}, vol.~52, pp. 167--174, 2019.

\bibitem{8933554}
Y.~{Li}, Y.~{Jiang}, D.~{Tian}, L.~{Hu}, H.~{Lu}, and Z.~{Yuan}, ``Ai-enabled
  emotion communication,'' \emph{IEEE Network}, vol.~33, no.~6, pp. 15--21,
  2019.

\bibitem{aloufi2020paralinguistic}
R.~Aloufi, H.~Haddadi, and D.~Boyle, ``Paralinguistic privacy protection at the
  edge,'' \emph{arXiv preprint arXiv:2011.02930}, 2020.

\bibitem{latif2017mobile}
S.~Latif, R.~Rana, J.~Qadir, A.~Ali, M.~A. Imran, and M.~S. Younis, ``Mobile
  health in the developing world: Review of literature and lessons from a case
  study,'' \emph{IEEE Access}, vol.~5, pp. 11\,540--11\,556, 2017.

\bibitem{taleb2017multi}
T.~Taleb, K.~Samdanis, B.~Mada, H.~Flinck, S.~Dutta, and D.~Sabella, ``On
  multi-access edge computing: A survey of the emerging 5g network edge cloud
  architecture and orchestration,'' \emph{IEEE Communications Surveys \&
  Tutorials}, vol.~19, no.~3, pp. 1657--1681, 2017.

\bibitem{liu2020toward}
Y.~Liu, M.~Peng, G.~Shou, Y.~Chen, and S.~Chen, ``Toward edge intelligence:
  multiaccess edge computing for 5g and internet of things,'' \emph{IEEE
  Internet of Things Journal}, vol.~7, no.~8, pp. 6722--6747, 2020.

\bibitem{ganin2015unsupervised}
Y.~Ganin and V.~Lempitsky, ``Unsupervised domain adaptation by
  backpropagation,'' in \emph{International conference on machine
  learning}.\hskip 1em plus 0.5em minus 0.4em\relax PMLR, 2015, pp. 1180--1189.

\bibitem{busso2008iemocap}
C.~Busso, M.~Bulut, C.-C. Lee, A.~Kazemzadeh, E.~Mower, S.~Kim, J.~N. Chang,
  S.~Lee, and S.~S. Narayanan, ``{EMOCAP}: Interactive emotional dyadic motion
  capture database,'' \emph{Language resources and evaluation}, vol.~42, no.~4,
  p. 335, 2008.

\bibitem{burkhardt2005database}
F.~Burkhardt, A.~Paeschke, M.~Rolfes, W.~F. Sendlmeier, and B.~Weiss, ``A
  database of german emotional speech,'' in \emph{Ninth European Conference on
  Speech Communication and Technology}, 2005.

\bibitem{costantini2014emovo}
G.~Costantini, I.~Iaderola, A.~Paoloni, and M.~Todisco, ``Emovo corpus: an
  italian emotional speech database,'' in \emph{International Conference on
  Language Resources and Evaluation (LREC 2014)}.\hskip 1em plus 0.5em minus
  0.4em\relax European Language Resources Association (ELRA), 2014, pp.
  3501--3504.

\bibitem{kaya2016robust}
H.~Kaya, A.~A. Karpov, and A.~A. Salah, ``Robust acoustic emotion recognition
  based on cascaded normalization and extreme learning machines,'' in
  \emph{International Symposium on Neural Networks}.\hskip 1em plus 0.5em minus
  0.4em\relax Springer, 2016, pp. 115--123.

\bibitem{latif2019unsupervised}
S.~Latif, J.~Qadir, and M.~Bilal, ``Unsupervised adversarial domain adaptation
  for cross-lingual speech emotion recognition,'' in \emph{2019 8th
  International Conference on Affective Computing and Intelligent Interaction
  (ACII)}.\hskip 1em plus 0.5em minus 0.4em\relax IEEE, 2019, pp. 732--737.

\bibitem{latif2018transfer}
S.~Latif, R.~Rana, S.~Younis, J.~Qadir, and J.~Epps, ``Transfer learning for
  improving speech emotion classification accuracy,'' \emph{Interspeech 2018:
  Proceedings}, pp. 257--261, 2018.

\bibitem{latif2018variational}
S.~Latif, R.~Rana, J.~Qadir, and J.~Epps, ``Variational autoencoders for
  learning latent representations of speech emotion: a preliminary study,''
  \emph{Interspeech 2018: Proceedings}, pp. 3107--3111, 2018.

\bibitem{schuller2013interspeech}
B.~Schuller, S.~Steidl, A.~Batliner, A.~Vinciarelli, K.~Scherer, F.~Ringeval,
  M.~Chetouani, F.~Weninger, F.~Eyben, E.~Marchi \emph{et~al.}, ``The
  interspeech 2013 computational paralinguistics challenge: Social signals,
  conflict, emotion, autism,'' in \emph{Proceedings INTERSPEECH 2013, 14th
  Annual Conference of the International Speech Communication Association,
  Lyon, France}, 2013.

\bibitem{schuller2014interspeech}
B.~Schuller, S.~Steidl, A.~Batliner, J.~Epps, F.~Eyben, F.~Ringeval, E.~Marchi,
  and Y.~Zhang, ``The interspeech 2014 computational paralinguistics challenge:
  Cognitive \& physical load, multitasking,'' in \emph{Proceedings INTERSPEECH
  2014, 15th Annual Conference of the International Speech Communication
  Association, Singapore}, 2014.

\bibitem{weninger2013acoustics}
F.~Weninger, F.~Eyben, B.~W. Schuller, M.~Mortillaro, and K.~R. Scherer, ``On
  the acoustics of emotion in audio: what speech, music, and sound have in
  common,'' \emph{Frontiers in psychology}, vol.~4, p. 292, 2013.

\bibitem{eyben2013recent}
F.~Eyben, F.~Weninger, F.~Gross, and B.~Schuller, ``Recent developments in
  opensmile, the munich open-source multimedia feature extractor,'' in
  \emph{Proceedings of the 21st ACM international conference on Multimedia},
  2013, pp. 835--838.

\bibitem{xu2015empirical}
B.~Xu, N.~Wang, T.~Chen, and M.~Li, ``Empirical evaluation of rectified
  activations in convolutional network,'' \emph{arXiv preprint
  arXiv:1505.00853}, 2015.

\bibitem{jaiswal2019privacy}
M.~Jaiswal and E.~M. Provost, ``Privacy enhanced multimodal neural
  representations for emotion recognition,'' \emph{arXiv preprint
  arXiv:1910.13212}, 2019.

\bibitem{pearson1901liii}
K.~Pearson, ``{LIII}. on lines and planes of closest fit to systems of points
  in space,'' \emph{The London, Edinburgh, and Dublin Philosophical Magazine
  and Journal of Science}, vol.~2, no.~11, pp. 559--572, 1901.

\end{thebibliography}




\end{document}